\documentclass[12pt,preprint]{aastex}

\def\lesssim{\mathrel{\hbox{\rlap{\hbox{\lower4pt\hbox{$\sim$}}}\hbox{$<$}}}}
\def\gtrsim{\mathrel{\hbox{\rlap{\hbox{\lower4pt\hbox{$\sim$}}}\hbox{$>$}}}}

\def\delvscale{\left({\Delta v \over \hbox{10~km~s$^{-1}$}}\right)}
\def\ergs{\hbox{erg~s$^{-1}$}}
\def\kms{\hbox{km~s$^{-1}$}}

\slugcomment{to appear in ApJ, 1 November 2004}
\shorttitle{ULXs as IMBHs}
\shortauthors{Krolik}
\usepackage{graphicx}

\begin{document}

\title{Are Ultra-Luminous X-ray Sources Intermediate Mass Black Holes
Accreting from Molecular Clouds?}

\author{Julian H. Krolik}
\affil{Physics and Astronomy Department\\
Johns Hopkins University\\ 
Baltimore, MD 21218}
\and
\affil{Institute of Astronomy\\
Cambridge University\\
Cambridge UK}

\email{jhk@pha.jhu.edu}

\begin{abstract}

   The origin and nature of Ultra-Luminous X-ray sources (ULXs) is a contentious
and controversial topic.  There are ongoing debates about the masses of the objects
responsible, their sources of mass for accretion, and their relation to stellar
populations in galaxies.  A new picture of these objects is proposed in which
they are intermediate-mass black holes confined to the disks of their host
galaxies and accreting from the interstellar medium.  They are then preferentially
found in or near molecular clouds.  This model correctly predicts the shape of the
observed luminosity function and requires only a very small fraction of
the baryonic mass of a galaxy to be in the form of intermediate-mass black holes.
Because the X-rays they produce strongly heat nearby interstellar
gas and because they move relatively rapidly in and out of dense regions,
ULXs are predicted to have brief episodes of high luminosity,
perhaps $\sim 10^5$~yr in duration, but they may recur many times.

\end{abstract}

\section{Introduction}

    Ultra-Luminous X-ray Sources (ULXs) are a class of X-ray source with apparent
luminosities in the range $10^{39}$ -- $10^{41}$\ergs seen in many nearby galaxies
(see, e.g., the reviews by van der Marel 2003 and Miller \& Colbert
2004).  Because these luminosities, if emitted isotropically,
would be super-Eddington coming from ordinary stellar black holes with
masses $\sim 10M_{\odot}$, their nature is the subject of much controversy.
Some believe that they are black holes of mass $\sim 10^2$--$10^3 M_{\odot}$
(Colbert \& Mushotzky 1999, Miller \& Hamilton 2002, Cropper et al. 2004).
Others argue that they are stellar mass black holes whose radiation
only appears to be super-Eddington because it is strongly anisotropic,
due either to disk transfer effects (King et
al. 2001), or relativistic beaming (K\"ording et al. 2002).
Still others suggest that they are stellar mass black holes able to
radiate at super-Eddington rates by means of a magnetic confinement mechanism
(Begelman 2002) or non-conservative highly super-Eddington accretion (King 2004).

     One clue to their origin is that they are often found close
to regions of active star formation (Matsushita et al. 2000, Roberts et al.
2002, Zezas et al. 2002, Kilgard et al. 2002, Roberts et al. 2004).
In fact, there are sometimes indications (either
through associated molecular line emission: Matsushita et al. 2000, or
large column density X-ray absorption: Zezas et al. 2002) that
these X-ray sources are within dense molecular clouds.  Initial claims
(e.g., Angelini et al. 2001) that there were numerous
ultra-luminous X-ray sources in early-type galaxies have been
reduced by Irwin et al. (2003) on the ground that all but two of the
brighter sources are chance
superpositions (but see Jeltema et al. 2003 for a better-supported
example).  In this paper we therefore explore the possibility that
ULXs (or at least the more numerous
variety in star-forming regions) may be intermediate
mass black holes that, when placed in dense molecular clouds,
accrete mass at a high enough rate to become X-ray luminous.

\section{The Velocity Dispersion}

\subsection{Centrality of the Black Hole Velocity Dispersion}

   The luminosity of a black hole accreting from the
interstellar medium depends on its
mass, the ambient gas density, and its speed relative to the surrounding gas:
\begin{equation}\label{bondiest}
L \sim 4 \times 10^{31} (\eta/0.1){\cal M}^2 n \delvscale^{-3}\ergs ,
\end{equation}
where $\eta$ is the radiative efficiency in rest-mass units,
${\cal M}$ is the mass of the black hole in solar masses, and $n$
is the density of atoms in cm$^{-3}$.  To make this estimate, we have used the
classical Bondi theory in the limit that the relative speed between
the black hole and the surrounding gas is larger than the thermal
speed of the gas.  Although stellar-mass black holes accreting from
regions of diffuse gas will never come close to radiating the luminosities
observed from ULXs (King et al. 2001, Agol \& Kamionkowski 2002), black holes
with ${\cal M} \gtrsim 300$ accreting from dense molecular clouds can
easily achieve luminosities $\gtrsim 10^{39}\ergs$.

     That is, this luminosity level can be achieved under two provisos:
that $\Delta v$ is not $\gg 10$~km~s$^{-1}$, and
the inner part of the accretion flow is disk-like so that relativistic
radiative efficiency can be achieved (see \S 3).  As equation~\ref{bondiest}
makes clear, the accretion luminosity is then extremely sensitive to
$\Delta v$.

      In fact, the velocity dispersion of the black holes also enters in
a second important way, governing the fraction of the population of black
holes found within a molecular cloud at any given moment.
Agol \& Kamionkowski (2002) estimated that the volume filling fraction of
dense molecular clouds near the Sun is $\sim 10^{-3}$ (for this purpose,
``dense" is defined as density greater than $n_{min} = 100$~cm$^{-3}$).
If we use their equation 8 scaled to the
data from galaxies with numerous ULXs such as M82 (Walter et al. 2002) or
the Cartwheel galaxy NGC 4038/9 (Wilson et al. 2003), the volume filling
fraction $f_{mol}$ in the galactic midplane in these galaxies
could be $\sim 50$ times larger (where there are probably factors of
several uncertainty owing to uncertainty in the CO/H$_2$ conversion ratio).
The scale height of molecular clouds in the Galaxy is
$\simeq 75$~pc, corresponding to a velocity dispersion $\sigma_{mol,z}
\simeq 10$~km~s$^{-1}$; if the black holes have a vertical velocity dispersion
$\sigma_{bh,z} > \sigma_{mol}$, the probability that any particular black hole
is found within a molecular cloud becomes $f_{mol}\sigma_{mol,z}/\sigma_{bh,z}$.

      Moreover, because the accretion rate is so sensitive to relative
speed, the fraction of black holes resident in a molecular cloud, yet
capable of accreting at a significant rate, is further reduced by the fraction
of phase space in which the velocity of the black hole is matched closely
to the orbital velocity of the molecular cloud.  When $\sigma_{bh} \gg
10$~km~s$^{-1}$, this fraction is $\sim (\sigma_{bh}/10$~km~s$^{-1})^{-3}$.

      Thus, for all these reasons, the critical factor in determining
the efficiency of intermediate mass black holes accreting from the
interstellar medium is their distribution of velocities relative to the
orbital velocity of molecular clouds.

\subsection{Initial Conditions for the Black Hole Velocity Dispersion}

    As with many problems, there are issues of both ``nature" and
``nurture".  We start with the initial state of the black hole
orbits, but these orbits are likely to evolve over a Hubble time.

\subsubsection{Black holes formed in mini-haloes}

    In one popular scenario for the origin of intermediate mass black holes,
they are formed at high redshift in dark matter ``mini-haloes" (Madau \&
Rees 2001, Schneider et al. 2002) when heavy element abundances are essentially
nil, so that the stellar initial mass function strongly favors stars with
${\cal M} \gtrsim 100$ (Bromm et al. 1999, Abel et al. 2000).  It is possible
that as much as $\sim 10^{-4}$ of the baryonic mass could end up in
intermediate mass black holes, arriving there through this pathway
(Madau \& Rees 2001).  These mini-haloes
then deliver the black holes to larger galaxies in mergers.

    Because the black holes are as collisionless as the dark matter, one would expect
their orbits in post-merger galaxies to be similar to the dark matter's,
i.e., appropriate to a galactic halo population (Schneider et al. 2002).
Some black holes may carry with them remnants of their dark-matter haloes,
and the added mass would enhance dynamical friction, but those that settle
deep into the post-merger galaxy are likely to lose these haloes by
tidal stripping (Islam et al. 2003). 
The typical relative speed between an intermediate
mass black hole of this sort and molecular clouds in the galactic disk would
then be $\sim 200$~\kms, far too large to permit accretion rates from the
interstellar medium capable of
sustaining an ultra-luminous X-ray source.  Only that small fraction whose
velocity fortuitously coincides with a molecular cloud's would be able to accrete
at a substantial rate.   Moreover, when considerations of accretion
and black hole binary dynamics are added to black hole merger rates,
the number of extra-nuclear black holes per galaxy acquired by this
mechanism is likely to be relatively small (Volonteri et al. 2003).

     Except for those black holes whose orbits take them either very close to
the center of the galaxy (within $\sim 10$~pc: Madau \& Rees 2001) or nearly
parallel to the disk plane (so that the time-averaged mean density of matter
encountered is $\gtrsim 1 M_{\odot}$~pc$^{-3}$), dynamical friction will do
little to black holes on such orbits even over a Hubble time.  Thus, for
the most part, the ``nurture"
side has little effect on the character of black hole orbits if they are
formed in this way.

\subsubsection{Black holes formed late from primordial abundance gas}

      Formation in mini-haloes is not, however, the only conceivable pathway
to creation of intermediate mass black holes.  It is possible, for example, that
gas of primordial elemental abundance (i.e., $ <3\times 10^{-4} Z_{\odot}$ in the
estimate of Bromm et al. 2001) also finds its way directly into the
disks of galaxies.  If stars are then created from this gas, one would
still expect their initial mass function to favor high-mass stars.  A population of
intermediate-mass black holes with disk orbits would soon result.  As will be
shown in \S 3, if the intermediate-mass black holes are a disk population,
only a very small total mass is required in order to produce the observed ULXs.  The
question, then, is whether some small fraction of the Universe's baryons can
be kept at primordial abundance until relatively late (perhaps $z \lesssim 1$),
be captured at that point into mature disk galaxies, and undergo star formation
before significant chemical mixing occurs.

      Most of the gas in the Universe is first ``polluted"
when it is pulled into a mini-halo and exposed to star formation there.
However, we do not know whether this is true of {\it all}
the baryons.  Although the standard CDM power spectrum of density
fluctuations is ``blue" enough that most matter
should be accreted into small-mass structures early on, it
seems implausible that this should happen to absolutely all baryons.
Unfortunately, quantitative estimates of the completeness
of this process are very difficult, either because exceptions to baryon
capture are ruled out by construction (as in the Press-Schechter formalism)
or because of limitations in resolution (in the case of numerical simulations).
At the very least, there should be a high-wavenumber cut-off beyond which
mini-haloes cannot accrete gas because, for example, their potential
depths are less than the temperature of the smoothly-distributed
baryons.  We suggest, therefore, that a small fraction of the
baryons escape capture into mini-haloes at high redshift.

     It is also possible that some gas that {\it is} accreted onto
small structures at high-redshift is nonetheless preserved from contamination
with heavy elements.  Shock waves driven by the first stars in these structures
could eject gas from low-binding-energy mini-haloes without substantial mixing;
tidal encounters could remove gas from their exteriors, etc.

    Although parts of the intergalactic medium do contain heavy elements
(as seen, for example, in quasar absorption line systems), the absence of
stellar stirring may keep the IGM sufficiently
quiescent as to limit mixing processes (see, e.g., the discussion of the many
uncertainties in Madau et al. 2001).
The median of the [C,O/H] distribution
in Ly$\alpha$ absorption systems is $\sim 10^{-3}$ Solar, but there is a wide
range of abundances seen (Simcoe et al. 2004).  Indeed, Simcoe et al.
argue that only about half the intergalactic baryons have been enriched
to greater than $10^{-3.5}$ Solar abundance in C and O by $z \simeq 2.5$.

   On the basis of these (admittedly loose) arguments, we conclude that it
is at least a reasonable possibility to suppose that some primordial abundance
material can be injected into galactic disks at relatively late times.  The
black holes created from this gas,
unlike the ones created in mini-haloes, would be born with orbits in the
galactic disk.  Their initial velocity dispersion relative to the local
circular orbital velocity should therefore be as small as for other young
stars ($< 10\kms$: Wielen 1977).  The absence of primordial abundance stars
in the Galactic disk provides only a weak constraint on this mechanism
if the initial mass function was, in fact, strongly weighted toward high-mass
stars because in that case only very few low-mass long-lived stars would
have formed from this gas.

\subsubsection{Black holes formed from stellar mergers}

     Another way intermediate mass black holes could be created inside a galactic
disk is through runaway merger of massive stars in a young cluster (Portegies
Zwaart et al. 2004).  This mechanism does not depend on having low
heavy-element abundances in the star-forming gas, so it can occur at any time
in the life of the galaxy.  Once created, these black holes, too, should
have random velocities characteristic of young stars or even smaller.

\subsection{Time-evolution of the Black Hole Velocity Dispersion}

   The mechanisms that control the evolution of stellar random motions
are not entirely understood, but there are two strong candidates:
encounters with molecular clouds (Spitzer \& Schwarzschild 1951, 1953) and
irregularities in the galactic potential (Barbanis \& Woltjer 1967, Carlberg
\& Sellwood 1985; Binney \& Lacey 1988, Jenkins \& Binney 1990).  Both
should act just as effectively on intermediate mass black holes as on stars
because, even if their masses are $\sim 10^3 M_{\odot}$, these black holes
are still much less massive than either giant molecular clouds or the mass
contributing to galactic potential fluctuations.

   On that basis, the observed velocity dispersion of old normal
stars provides an approximate upper bound on the dispersion of
intermediate mass black holes.  For example, in our Galaxy, the
vertical dispersion of the oldest
stellar population is $\simeq 36\kms$, while its radial dispersion is
$\simeq 62\kms$ (Reid et al. 2002).  However, the dispersions of
intermediate mass black holes may be significantly smaller for two
reasons: if they were created relatively recently (by the merger-in-clusters
mechanism, for example), or if dynamical friction against comparatively
slowly-moving young ordinary stars diminishes their speed relative
to the local standard-of-rest.

    To estimate the magnitude of these mechanisms, we follow Jenkins \&
Binney (1990) and Binney \& Lacey (1988):  We describe the evolution of
the black hole distribution function in terms of a Fokker-Planck
equation and use their formalism for estimating the magnitudes of
the coefficients in that equation.  The basic equation is then
\begin{equation}
{\partial f \over \partial t} = - \nabla_E \left(\vec g f \right) +
 (1/2)\nabla_E \left(\cdot {\mathbf D} \cdot \nabla_E f\right),
\end{equation}
where the distribution function is with respect to random kinetic
energy in the vertical and radial directions, $\nabla_E$ is the gradient
with respect to kinetic energy associated with each axis,
$\vec g = \langle \dot E_R \rangle \hat R + \langle \dot E_z \rangle \hat z$, and
${\mathbf D}$ is the $2 \times 2$ matrix of diffusion coefficients.

    Dynamical friction accounts for $\vec g$:
\begin{equation}
g_i = -4\pi \ln \Lambda G^2 M_{\odot}\rho {\cal M} \chi /v_i.
\end{equation}
Here $\rho$ is the smoothed density of stars, $v_i = \sqrt{2E_i}$ is the
black hole's random speed in the $i$th direction, and
$$\chi \equiv  \hbox{erf}(X) - 2X \exp(-X^2)$$
for $X = |v_i|/(\sqrt 2 \sigma_*)$, where $\sigma_*$ is the stellar
velocity dispersion.  Evaluating $\vec g$ for typical values gives
\begin{equation}
g_i = -2.3 \times 10^{-6} ({\cal M}/1000) \left({v_i \over 10 \kms}\right)^{-1}
    \left({\rho \over 0.1 M_{\odot}\hbox{pc}^{-3}}\right)\chi
    \hbox{~cm$^2$~s$^{-3}$}.
\end{equation}
We choose a stellar density of $0.1 M_{\odot}$~pc$^{-3}$ as our fiducial
value because it is the local density near the Sun (Merrifield \& Binney 1999).

    In the Jenkins \& Binney (1990) formulation, all four of the components
in the diffusion coefficient matrix have contributions from scattering
with molecular clouds, but potential fluctuations contribute only to the
$R$--$R$ component.  According to their estimate, the characteristic scale
for all contributions is set by
\begin{equation}
C \equiv {8 \over \pi} \ln \Lambda G^2 \mu_c M_c \nu_z
  = 6 \times 10^6 \left({\mu_c \over 2 M_{\odot}\hbox{~pc$^{-2}$}}\right)
     \hbox{cm$^4$~s$^{-5}$} ,
\end{equation}
where $\mu_c$ is the mean surface density of molecular clouds, $M_c$ is the
mass-weighted mean cloud mass, and $\nu_z$ is the frequency of harmonic
motion in the vertical direction.  In the case of fluctuations in the potential,
$D_{RR} \simeq \beta J_{1}^2({\cal K}a)C$, where $\beta \simeq 90$ is a coefficient
determined by fitting to data, $J_1$ is the first-order Bessel function,
${\cal K}$ essentially defines the wavenumber of the disturbance, and
$a=\sqrt{2}v_R/\kappa$ is the amplitude of epicyclic motion for
radial epicyclic frequency $\kappa$.

   The relative importance of diffusion in velocity space and dynamical
friction can be gauged by comparing the characteristic rates with which
they alter the distribution function, $t_{fric,i}^{-1} \sim |g_i|/v_i^2$
and $t_{diff,i}^{-1} \sim D_{ii}/v_i^4$:
\begin{equation}
{t_{diff,i} \over t_{fric,i}} \sim 
    1 \left({\rho \over 0.1 M_{\odot}\hbox{pc$^{-3}$}}\right)
       \left({{\cal M} \over 1000}\right)
      \chi \left({\mu_c \over 2 M_{\odot}\hbox{pc$^{-2}$}}\right)^{-1}
      \cases{ v_{10}^{-1} & $i=R$ \cr
              0.25 v_{10} & $i=z$ \\}
\end{equation}
The dependence on $v_{10}$ reflects the assumption that scattering by potential
fluctuations dominates $D_{RR}$.  When the scaling quantities have their fiducial
values, the two rates are comparable.  Because diffusion
by potential fluctuations increases in strength as the random speed grows,
whereas scattering by clouds is independent of the random speed, dynamical
friction may be better able to restrain the growth of vertical motions than
horizontal motions.

    Unfortunately, because there remain quantitative problems
using any of these prescriptions to predict the actual stellar dispersions
in the Solar neighborhood (G. Gilmore, private communication), these scalings
should not be taken as definitive.  We are thus left with the conclusion
that the random speeds of intermediate mass black holes injected
early into a galactic disk are unlikely to be greater than those of
old stars ($\simeq 40$--60\kms), and it is possible, particularly
for those with masses $\gtrsim 1000 M_{\odot}$ found in regions
of relatively high stellar density, that their random speeds may be
rather closer to those of young stars ($\simeq 10$\kms).

\section{Population Estimates}

      As remarked above, the luminosity function of black holes in
a galaxy accreting from its interstellar medium depends on their mass distribution,
their velocity distribution, and the distribution of gas density.  To
predict the luminosity function, we employ a formalism very similar to that
of Agol \& Kamionkowski (2002), i.e.,
\begin{equation}
{dN \over dL} = \int \, dM \, \int \, d^3v \, \int \, d^3x \, \int \, dn_g
                n_{BH}(\vec x) f(\vec v) {dn_{BH} \over dM} p(n_g,\vec x)
                \delta\left[L - 4\pi n_g \bar{m} c^2 \eta(GM)^2/v^3\right],
\end{equation}
where $\vec x$ is position in the galaxy, $n_g$ is the gas density,
$n_{BH}$ is the density of intermediate mass black holes, $f(\vec v)$ is
the velocity distribution function of the black holes normalized to unity,
$dn_{BH}/dM$ is the mass function of the black holes similarly normalized
to unity, and $p(n_g,\vec x)$ is the probability of finding gas of density
$n_g$ at location $\vec x$.  For this toy-model, we suppose that the
vertical distribution of both the black hole density and the gas density
are Gaussians, but with scale-heights determined by their differing
velocity dispersions.  In the interest of simplicity, we likewise
suppose that the black hole velocity distribution is an isotropic
Gaussian.  Following Agol \& Kamionkowski, we suppose that $p(n_g,\vec x)
\propto n_g^{-\beta}$ from $n_{min}$ to $n_{max}$; they take $\beta \simeq 2.8$,
$n_{min} = 100$~cm$^{-3}$ and $n_{max} = 10^5$~cm$^{-3}$.  Lastly,
for no particularly good reason other than ease of computation, we
write $dn_{BH}/dM \propto M^{-\alpha}$ from $M_{min}$ to $M_{max}$, where we
imagine that $1 < \alpha < 3$, $M_{min} \sim 100M_{\odot}$,
and $M_{max} \sim 10^4M_{\odot}$.  With those assumptions,
\begin{equation}\label{lumfunct}
{dN \over dL} = {2^{5-2\beta} (2\pi)^{9/2-2\beta} \over 3}{(\alpha - 1)(\beta-1) \over
                (\beta - 2)(3 - \alpha)} f_{mol} {h_g \over h}\left({M_{min} \over
                M_{max}}\right)^{\alpha - 1} {L_* \over L}{N_{tot} \over L}
                \exp\left[ -(1/2) (L_*/L)^{2/3}\right],
\end{equation}
where $h,h_g$ are the scale-heights of the black holes and the gas, respectively,
and we suppose $h_g < h$; $L_* = 4\pi n_{min} \bar{m}c^2 \eta (GM_{max})^2/
\sigma_{bh}^3$ is a characteristic luminosity, the luminosity of a maximum-mass
black hole accreting from a minimum density cloud; and $N_{tot}$ is the total
number of intermediate
mass black holes in the galaxy.  Strikingly, the predicted shape of $dN/dL$, i.e.,
$\propto L^{-2}$, is {\it independent} of $\beta$ and $\alpha$.  This result
follows from the coincidence that the Bondi accretion rate is $\propto v^{-3}$
while we have guessed that the black hole velocity distribution function is
quasi-thermal, i.e., $dn \propto \exp(-v^2/2\sigma^2) d(v^3)$.

  Rewriting equation~\ref{bondiest} in terms of the characteristic quantities,
we see that
\begin{equation}\label{lstar}
L_* = 6 \times 10^{39} \left({\eta \over 0.1}\right)
                       \left({{\cal M}_{max} \over 10^4}\right)^2 \left({n_{min}
      \over 100\hbox{cm$^{-3}$}}\right)\left({\sigma_{bh} \over 40\kms}\right)^{-3}
      \hbox{erg~s$^{-1}$}.
\end{equation}
Making use of this lower cut-off to the luminosity function, we find that
the fraction of all intermediate mass black holes found in regions where
the Bondi accretion rate is great enough to support ULX-level luminosity
(defined here as $L > L_*$) is
$\sim f_{mol}(h_g/h)(M_{min}/M_{max})^{\alpha -1}$.  A change in the
(somewhat arbitrary) definition of $n_{min}$ would be reflected in a
change in the radiating fraction through its implications for $f_{mol}$.

    As already discussed, $f_{mol}$ could rise to as great as $\sim 0.05$
where star formation is occurring at a rapid rate.  If $h_g/h\simeq 1/4$,
as implied by our guesses about the velocity dispersions, the total number
of intermediate mass black holes in a galaxy displaying $\sim 10$ ULXs
would be $N_{BH} \sim 800 (M_{max}/M_{min})^{\alpha -1}$.  This scaling assumes
$1 < \alpha < 3$; the dependence on $M_{max}/M_{min}$ arises because the
lowest-mass black holes are likely
to dominate in number, but because the accretion rate is $\propto M^2$,
the highest-mass contribute disproportionately to the numbers of bright
objects.  The total mass in intermediate-mass black holes in a galaxy
is $\sim f_{mol}^{-1}(h/h_g) M_{max} N(L>L_*) F$, where $F = 1$ if $\alpha < 2$
or $(M_{max}/M_{min})^{\alpha - 1}$ if $\alpha > 2$.  For example,
if $\alpha = 2$ and $M_{max}/M_{min}=100$, the total mass in
intermediate-mass black holes is $\sim 4 \times 10^6 N(L>L_*) M_{\odot}$.
This would be a small fraction, indeed, of the total baryonic mass of the
galaxy.

    It is encouraging that the slope of the luminosity function predicted
in equation~\ref{lumfunct} (-1 for the cumulative function) is in quite
good agreement with the slope of the luminosity function above
$10^{39}$~erg~s$^{-1}$
found by Kilgard et al. (2002) in the Antennae galaxies.

  On the other hand, in order to produce the observed population of
ultra-luminous X-ray sources by means of that small fraction of intermediate
mass black holes formed in mini-haloes and capable of radiating brightly, the
total number of such black holes in a typical galaxy must be very large.
In that case, the bright fraction is only $\sim f_{mol}(\sigma_{mol}/\sigma_{bh})^4
\sim 10^{-6}$, and to produce 10 ULXs in a galaxy would likely require more
than $2 \times 10^9 M_{\odot}$
in intermediate mass black holes.  Thus, if ULXs are indeed intermediate-mass
black holes accreting from dense gas in the host galaxy's interstellar medium,
possessing disk-like kinematics is essential: the total mass required by
a bulge- or halo-like population would be implausibly large.  The only
escape from this argument comes from the possibility that black holes of this variety
might accrete, not from the general interstellar medium, but from baryonic
matter they brought with them from the mini-halo in which they were created.
If this gas is substantially more tightly bound to them than
the dark matter, it would be better able to resist tidal stripping, but
fairly extreme assumptions about this process are required to create
large numbers of ultra-luminous X-ray sources (Islam et al. 2003).

\section{Impact on the Molecular Cloud and the Duty Cycle of Activity}

  Such a large ionizing luminosity inflicted upon a nearby molecular cloud
would be sure to change its state dramatically.  For example, Neufeld et al.
(1994) described the impact of X-rays impinging on nearby molecular gas
in terms of the parameter $F/(N^{0.9}p)$, where $F$ is the X-ray flux,
$N$ the gas's column density, and $p$ its pressure.  They found that when
this parameter exceeds $10^{-13}$~cgs, molecules are destroyed.  In terms
of fiducial ULX numbers, the Neufeld et al. parameter in the circumstances
of interest here is $\sim 10^{-10} L_{40}(r/10$~pc$)^{-3}(n/100$~cm$^{-3})^{-2}$~cgs,
a thousand times greater than the critical value.

   The surrounding gas should therefore be strongly photoionized and heated.
In photoionization equilibrium, its state can be conveniently
described in terms of the pressure-ratio photoionization parameter
$\Xi \equiv L_{ion}/(4\pi r^2 n_H kT)$.  When the density refers entirely
to spherically-accreting gas, the ionization parameter can be rewritten
as
\begin{equation}
\Xi = 3.9 c\max(v_r,c_s)/c_s^2,
\end{equation}
where $v_r$ is the radial infall speed, $c_s$ is the gas's sound
speed, and a fully-ionized H-dominated chemical composition has
been assumed.  Values of $\Xi > 1$ are to be expected out to
distances where $v_r/c_s \sim c_s/c$, and where $\Xi \gg 1$, the
gas will be hot and highly ionized.

    Photoionization equilibrium is, in fact, a likely condition.  Comparing
the ionization timescale for H with the accretion timescale $t_{acc} \equiv
r_A/v$ yields the
ratio
\begin{equation}
{t_{ion,H} \over t_{acc}} = \tau_H^{-1} \left({L_{ion}/L \over \eta}\right)
                   \left({\langle\epsilon\rangle \over \bar{m}c^2}\right),
\end{equation}
where $\tau_H$ is the spectrally-averaged H-photoionization optical depth
across radius $r$ if all the gas were neutral, and $\langle \epsilon\rangle$
is the mean ionizing photon energy.  For our fiducial parameters,
$\tau_H \simeq 6 ({\cal M}/1000) \sigma_{bh,40}^{-2} n_{100}
(\langle \epsilon \rangle/I_H)^{-3}$, so ionization balance should be
easy to achieve in a flow-time provided the spectrum is not extremely
hard.

   If we assume a generic black hole spectrum, a sum of a thermal component
with a typical ULX temperature of 0.2 keV (Miller et al. 2004) and a
Comptonized power-law, the equilibrium (i.e., Compton) temperature for
$\Xi \gtrsim 30$ is $T_C \sim 10^7$~K.  Heating by photoionization
should be very rapid up to temperatures close to $T_C$;
XSTAR (Kallman \& Bautista 2001) calculations show that
\begin{equation}
t_{heat}/t_{acc} \sim 0.3 (T/T_C)^{2.5} n_{100}^{-1} r_{A,16}^{-1},
\end{equation}
where the accretion radius
$r_A = 8.3 \times 10^{15}({\cal M}/1000) \sigma_{bh,40}^{-2}$~cm.  Only
when a relatively small black hole is immersed in a relatively low
density region might the temperature fail to reach equilibrium in
an accretion time, but that is a combination of conditions yielding
only a small luminosity (see eqn.~\ref{lstar}).  Black hole masses
and interstellar densities high enough to provide a truly ULX
luminosity imply rapid thermal equilibration.

    When the temperature is raised close to $T_C$, the accretion flow is
severely disrupted.  Because the accretion rate is proportional to
$c_s^{-3}$ when $c_s > v$, the accretion rate would fall relative to our
nominal prediction by a factor $\sim 10^3$.  An immediate
drop in the luminosity by this factor would push $\Xi$ low enough to once
again permit a lower-temperature equilibrium, but in fact there are
likely to be significant time delays built into the system.

   As Agol \& Kamionkowski (2002) point out, density gradients within
the molecular cloud are likely to imprint a small net angular momentum
on the accreting matter.  At a radius small compared to the accretion
radius but large compared to the black hole, the flow must then flatten
and form a disk.  Their estimate applied in our context indicates that
this typically happens at
$r_{disk} \sim 2 \times 10^5 ({\cal M}/1000)^{2/3} \sigma_{bh,40}^{-10/3} r_g$.
The accretion radius $r_A \sim 10^8 \sigma_{bh,40}^{-2} r_g$, so $r_g \ll r_A$,
while at the same time $r_{disk} \gg r_g$,
justifying our assumption of relativistic radiative efficiency.

    The inflow time $t_{in}$ through a thin disk can be much longer than the
disk orbital time $t_{dyn}$, as it is limited by the slow process of angular
momentum transport.  In terms of the $\alpha$-model of Shakura \& Sunyaev
(1973), $t_{in} \sim \alpha^{-1} (r/h)^2 t_{dyn}$, where $\alpha$ is
the ratio of the mean vertically-integrated stress to the vertically-integrated
pressure and $h$ is the thickness of the disk.  Under the assumptions that
gas pressure dominates the vertical support of the disk and that the
disk is in inflow equilibrium at the Bondi accretion rate, we estimate
that on the scale of $r_{disk}$,
\begin{equation}
h/r \sim 5 \times 10^{-3}\left({\alpha \over 0.1}\right)^{-1/10}           
                         \left({{\cal M}\over 10^4}\right)^{1/15}
                         n_{100}^{1/5}\sigma_{bh,40}^{3/5}.
\end{equation}
We then find that
\begin{equation}
{t_{in}(r_{disk}) \over t_{acc}} \sim 900
                         \left({\alpha \over 0.1}\right)^{-4/5}           
                         \left({{\cal M}\over 10^4}\right)^{13/15}
                         n_{100}^{-2/5}\sigma_{bh,40}^{-16/5}.
\end{equation}

    If ambient conditions remained constant for long periods of time,
it might then be possible for the system to go through a limit-cycle
with characteristic time
\begin{equation}
t_{in}(r_{disk})\sim 6 \times 10^5 
                         \left({\alpha \over 0.1}\right)^{-4/5}           
                         \left({{\cal M}\over 10^4}\right)^{28/15}
                         n_{100}^{-2/5}\sigma_{bh,40}^{-31/5}\hbox{~yr}.
\end{equation}
While the ULX
luminosity is low, accretion could proceed rapidly and the mass of the
accretion disk surrounding the intermediate mass black hole could build;
once the luminosity rises, further addition of material to the disk would be
effectively halted.  Note that as the disk surface density $\Sigma$ grows,
the accretion rate through the disk rises $\propto \Sigma^{5/3}$, so
that the luminosity at early stages lags somewhat behind the build-up
of mass in the disk.

   During the bright phase of the ULX, the energy it injects into the
surrounding medium raises its temperature and pressure sharply.  An
expanding bubble should result, whose impact on the molecular cloud
should be substantial.  This should be true independent of the origin
of the ULX, provided only that its radiative output is roughly
isotropic.  There is, however, a significant complication due to
relative motion between the ULX and the molecular cloud: the
change in relative position during the bright phase is $\sim
25 (\alpha/0.1)^{-4/5} ({\cal M}/10^4)^{28/15} n_{100}^{-2/5}
\sigma_{bh,40}^{-26/5}$~pc, a distance comparable to or larger than
typical distinct dense regions in the interstellar medium.  The lifetime
in the bright phase might therefore be set by whichever is shorter,
the residence time of matter in the accretion disk or the passage time for
the black hole through a region of dense gas.
Consideration of the time-dependent interplay between X-ray heating
and motion of the black hole through a molecular cloud is an interesting topic
best deferred to subsequent work.

\section{Summary}

     We have shown that if intermediate-mass black holes are placed in
the denser parts of the interstellar media of disk galaxies, Bondi accretion
provides a great enough accretion rate to support the luminosities seen
in Ultra-Luminous X-ray sources.  The direct relation between ambient
density and accretion rate automatically leads to a correlation between
ULXs and regions of ongoing star-formation, as is observed.  Indeed,
to the extent that CO luminosity is a proxy for molecular gas
content, it may be an indicator of ULX activity (E. Agol, private
communication).  Cross-comparison of CO maps with X-ray images
for the nearer galaxies containing ULXs could then provide a test for this
model.  Note, however, that there is a rather strong selection bias
for finding ULXs immediately {\it outside} dense molecular cloud regions:
because their spectra are rather soft, they are very susceptible to
soft X-ray absorption, and column densities commonly found in molecular
clouds ($\sim 10^{22}$~cm$^{-2}$) block photons below $\simeq 2$~keV.
As noted at the end of the last section, the residence time of matter
in their accretion disks can be long enough for black holes
to move out into a clear region while still radiating brightly, having
acquired their accretion fuel when in a denser neighborhood.

      The central problem for this picture is whether sufficient intermediate-mass
black holes can be created with orbits that keep them where the dense gas
is located, i.e., close to the disk plane.  Because it is believed that
star-formation in primordial abundance gas is an especially efficient
path to the production of intermediate-mass black holes (Bromm et al. 1999;
Abel et al. 2000; Madau \& Rees 2001; Schneider et al. 2002), the key
question is whether enough gas free of heavy elements can be preserved
until the orientation of present-day disk galaxies
is set, brought into those disks, and then provoked into star-formation
before it mixes with the interstellar medium of those galaxies.  The
likelihood of a positive answer to this question is raised by the fact
that the amount required is a very small fraction of the total baryonic
mass of contemporary disk galaxies, but any answer to it is at present
highly speculative.  Alternatively, intermediate-mass black holes might
be formed as a disk population by runaway mergers in clusters (Portegies-Zwaart
et al. 2004).

      If intermediate-mass black holes are created within a galactic disk,
they are likely to have random velocities no greater, and possibly rather
smaller, than the old stars in the disk.  Employing very simple assumptions,
we predict that the shape of the luminosity function that
results from such relatively small random speeds should agree with that observed:
$dN/dL \propto L^{-2}$.  Particularly in galaxies with especially high
molecular gas content, the total mass in intermediate-mass black holes
necessary to create an active population with the numbers seen is quite
modest, possibly as little as $\sim 10^7 M_{\odot}$.

      However, placing such a strong X-ray source in an interstellar cloud
disrupts the very conditions that allow rapid accretion to take place.  Time
delays in the accretion disk that should form well within the nominal
Bondi accretion radius allow accretion to continue long enough to build up
a reservoir that can supply the X-ray source for $\sim 10^5$~yr
after it ionizes and heats its environment so thoroughly that further
Bondi accretion is shut off, or, depending on which happens first, it leaves
the region of dense gas in which it accumulated the mass in its
accompanying disk.  It is likely that individual sources
therefore undergo frequent cycling between bright and dim episodes.
The effects on surrounding molecular gas deserve
further detailed investigation.

\acknowledgements{JHK thanks Eric Agol, Andy Fabian, Gerry Gilmore, Martin
Haehnelt,
and Martin Rees for many helpful conversations, as well as the referee,
Piero Madau.  He is also grateful to the
Institute of Astronomy, Cambridge for its hospitality while this work
was completed, and to the
Raymond and Beverly Sackler Fund for support during his visit there.
His work is also partially supported by NSF grants AST-0205806 and AST-0313031.}

\end{document}